# Surface acoustic waves to monitor active THz metagrating based on $VO_2$

**G Y Karapetyan[1], M V Ochkurov[1] and V E Kaydashev[1]**

[1]Laboratory of Nanomaterials, Southern Federal University, Stachki 200/1, 344090 Rostov-on-Don, Russia

E-mail: vekaydashev@sfedu.ru

**Abstract.** We theoretically propose a new approach to *in-situ* monitor the altered states of pixels in $VO_2$ based active THz amplitude metagrating by using surface acoustic waves (SAWs) generated in $LiNbO_3$ substrate. A single broadband RF response of the SAW device consists of *N* narrowband frequency channels which code the feedback information on the current pixels states. This way the resistance alteration due to metal-to-insulator transition of all $VO_2$ based pixels of the active THz grating is monitored in advance even without incident radiation. The method provides important new options of monitoring a metasurface aging as well as incorrect switching related to technological defects. Using of the SAW device in all-electrically addressed $VO_2$ based binary grating with tunable period as well as advanced gradient type gratings are modeled.

## 1 Introduction

Metasurfaces pave a broad avenue for advanced flat optic tools and technologies due to their unique capabilities in the formation of previously inaccessible field distributions within the wavefront. Metasurfaces are extensively studied to build a basis for new methods of terahertz (THz) nondestructive testing, THz spectroscopy, THz imaging, microfluidic THz detection of biological species, thermal and THz monitoring, parallel optical processing, new generation wireless telecommunication systems and many more.

Passive metasurfaces consist of an array of metallic or dielectric antennas whose functions are predetermined during design and are not altered during work.[1–4] Active metasurfaces are based on phase change materials those properties are in-situ adjusted by electrical, optical, thermal and other means.[5–12] However, by now the properties of all meta-atoms in these devices are altered simultaneously by using a “global” gate voltage,[7,8] “global” electric bias applied to all meta-atoms.[6,9–12] Locally addressed single THz meta-atoms are still rare[13] because their fabrication requires time-consuming multistep lithography protocols. Hopefully, a few types of all-electric active metasurfaces could be designed even using a single optical lithography step without the need to design complicated topologies with multiple crossed wires to achieve complete electrical addressing. One of such devices, namely, diffraction metagrating is discussed in the present study.

Very recently a paradigm of all-optical data recognition using “traditional” Deep Neural Network (DNN) has been established and very soon it has been evolved to a Physics-Informed Neural Network (PINN) based approach.[14,15] In PINN a physical multi-layer diffracting optical medium is combined with mathematical Neural Network algorithm to avoid time-consuming network training which boosts the calculating performance. All-optical diffractive processor[16] proposed very recently evidenced a great potential of multi-layer diffractive network based on physical medium for all-optical computing. Active metasurfaces with built-in *in-situ* monitoring of all individual meta-atoms during operation or at least groups of meta-atoms (pixels) are highly desired as they are expected to considerably contribute to all-optical diffractive processors. Indeed, by only applying a voltage to "pixels" it can not be guaranteed that a THz metasurface is working correctly. Thus, feedback signals are highly desired.

Recently, by using SAW we experimentally examined an unusual metal-to-isolator transition (MIT) in $VO_2/TiO_2/LiNbO_3$ composite[17] that allowed us not just switch material between an isolator and metallic states but also to gradually tune its conductivity to any point of MIT hysteresis loop. Such unique property of the $VO_2/TiO_2/LiNbO_3$ composite opened a new way to build active metaurfaces of "gradient" type.

In this letter, we study a new approach of using surface acoustic waves (SAWs) generated in $LiNbO_3$ substrate to *in-situ* monitor the states of pixels in $VO_2$ based active THz metasurface. More specifically, we theoretically model a novel SAW-based device to simultaneously and independently monitor the THz reflection of all $N$ lines of active $VO_2$ based diffraction grating. A single broadband RF response of the SAW device consists of $N$ narrowband frequency channels which code the feedback information about the current state of each line (pixel). Moreover, all-electrically addressed $VO_2$ metasurface can be a binary amplitude grating as well as a sine amplitude grating.

The proposed approach is general and will be also useful for monitoring many other types of $VO_2$ based active THz metasurfaces.

## 2 Experimental procedure

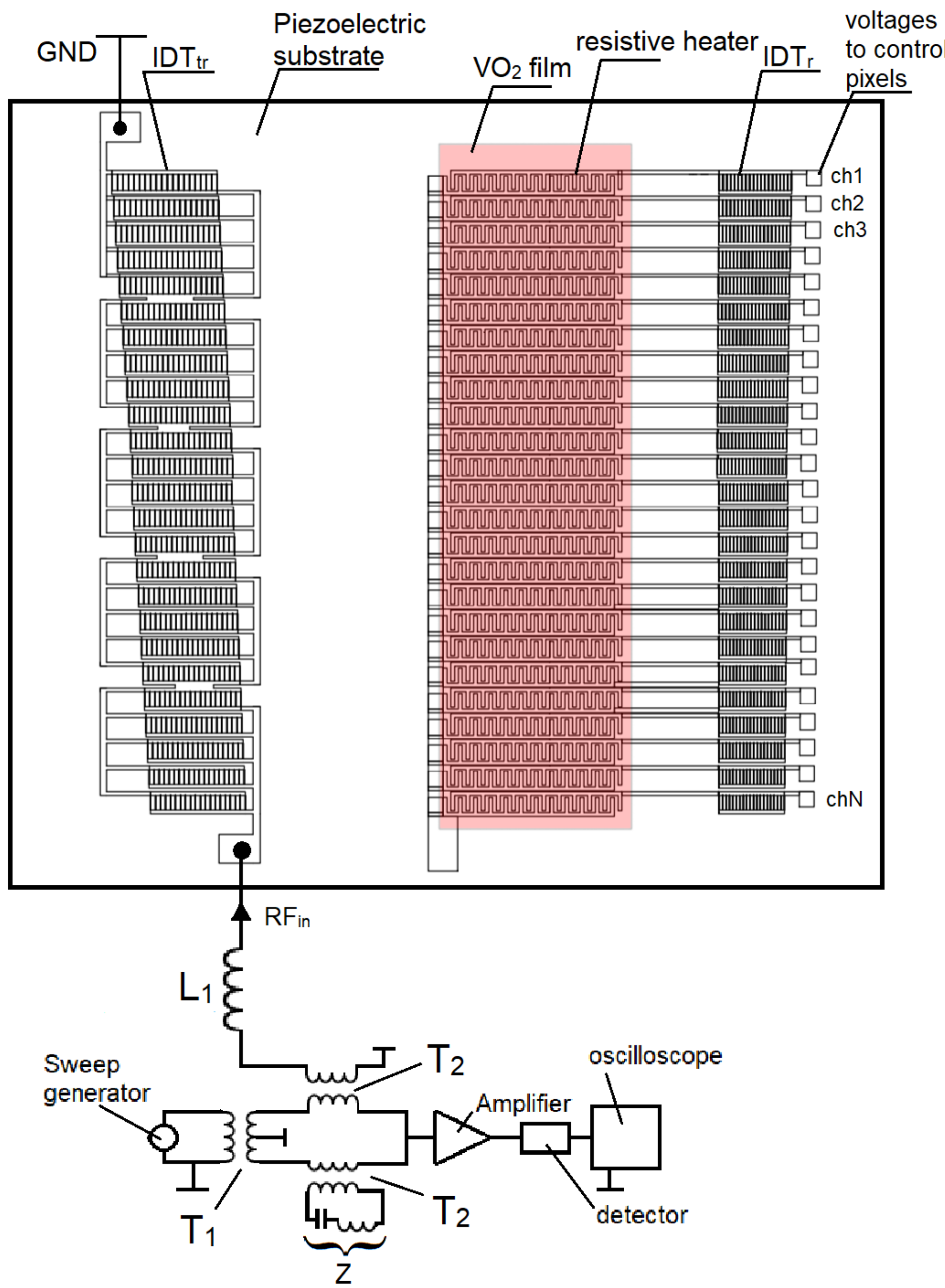

**Fig. 1** Schematics of multi-channel SAW device to monitor $VO_2$ based active THz metagrating

Transmitting-receiving (TR-IDTs) and reflecting (R-IDTs) inter-digital transducers shown in Figure 2 (a) and (b) respectively were specially designed to achieve 25 independent SAW frequency channels with non-overlapping narrow bandwidths of 2MHz. The corresponding characteristic $S_{21}$ frequency responses of the TR-IDT and R-IDT are shown in Figure 2 (c) and (d). S21 frequency response of a SAW reflected back from R-IDT is shown in Figure 2(e).

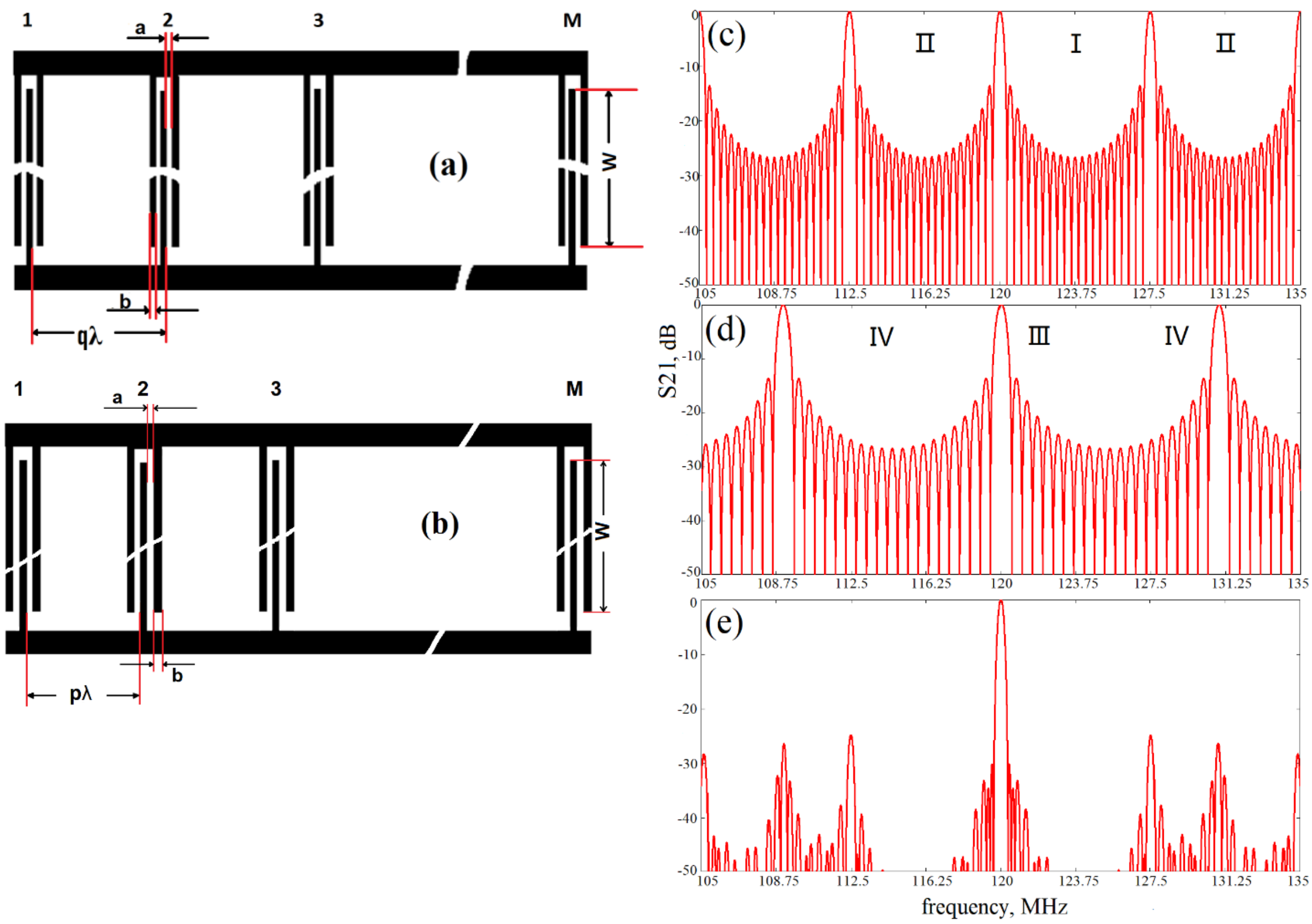

**Fig. 2** Geometry of a transmitting-receiving (TR-IDT) (a) and reflecting (R-IDT) (b) inter-digital transducers with a=b=8.3μm, w=1280μm. Frequency response of TR-IDT (c) and R-IDT (d), correspondingly. Frequency response of a SAW reflected back from R-IDT (e). IDTs consist of 20 single wave sections which are positioned at distance of $16\lambda$ (q=16) and $11\lambda$ (p=11) from each other for transmitting-receiving and reflecting IDTs, correspondingly. Note that IDT digits` positions are chosen at central SAW frequency for each channel. Aperture of IDTs is 225 $\mu$m and the acoustic channel length is $2060\lambda$.

The frequency position of the band I of the TR-IDT in Figure 2 (c) coincides with position of band III of the R-IDT in Figure 2 (d) while the frequency bands II in Figure 2 (c) and IV Figure 2 (d) do not overlap. Thus, the reflected SAW shown in Figure 2 (e) reveals the frequency response with a single pronounced band I. Normalized frequency responses of single TR-IDT and R-IDT are expressed by [18]

$$H_{\mathrm{TR-IDT}} = \frac{1}{M}\sqrt{\left[\sum_{k=0}^{M}\left(\cos\left(\frac{2\pi qkf}{f_0}\right)\right)\right]^2 + \left[\sum_{k=0}^{M}\left(\sin\left(\frac{2\pi qkf}{f_0}\right)\right)\right]^2}$$

$$H_{\mathrm{R-IDT}} = \frac{1}{M}\sqrt{\left[\sum_{k=0}^{M}\left(\cos\left(\frac{2\pi pkf}{f_0}\right)\right)\right]^2 + \left[\sum_{k=0}^{M}\left(\sin\left(\frac{2\pi pkf}{f_0}\right)\right)\right]^2} \qquad (1)$$

where M is a number of sections in IDTs, M=20, $q$ is a number of SAW periods corresponding to the distance between adjacent active digits of IDT as shown in Figure 2: $q$=16 for transmitting-receiving TR-IDT as shown in Figure 2(a) and $q$=11 for reflecting R-IDT (see Figure 2(b)).
The frequency band of IDT is roughly equal to $f_0/(Mq)$. Thus, when the difference between central frequencies of the neighboring channels $\Delta f = f_i - f_{i+1}$ is larger than $f_0/M$, then their frequency bands are not overlapped and the inter-channel influence is minor. According to equation (1) the positions of frequency bands for TR-IDT and R-IDT do not coincide when the $q\cdot\lambda$ and $p\cdot\lambda$ values, i.e. the distances between adjacent sections in IDTs, are not multiples of each other.

The SAW delay line is described with admittance matrix $Y$ as:[19]

$$\begin{bmatrix} Y_{11} & Y_{12} \\ Y_{12} & Y_{22} \end{bmatrix},$$

where $Y_{11}=Y_{tr}$ is an admittance of transmitting-receiving TR-IDT, $Y_{22}=Y_r$ – is an admittance of receiving R-IDT, $Y_{tr}=Ga_{tr}+j(Ba_{tr}+\omega C_{tr})$, $Y_r=Ga_r+j(Ba_r+\omega C_r)$, $Y_{12}=\sqrt{Ga_{tr}\cdot Ga_r}\cdot e^{j\omega T}$, $T=2L/V_{SAW}$, $L$ – the acoustic channel length, i.e. the distance between TR-IDT and R-IDT. $Ga_{tr}$ and $Ga_r$ are active (real) and $Ba_{tr}$ and $Ba_r$ reactive (imaginary) parts of admittance for transmitting-receiving and reflecting IDTs, respectively. $C_{Ttr}$ and $C_{Tr}$ are static capacities of TR-IDT and R-IDT, correspondingly

$$Ba \approx 8\mathrm{f}_0 k_{eff}^2 \mathrm{C}_T N \frac{\sin(2\mathrm{X})-2\mathrm{X}}{2\mathrm{X}^2} \tag{2}$$

$$Ga \approx 8\mathrm{f}_0 k_{eff}^2 \mathrm{C}_T N \left(\frac{\sin X}{X}\right)^2 \tag{3}$$

where $k_{eff}$ is a electro-mechanical coupling coefficient and $X=\pi N\frac{f-f_0}{f_0}$.

An important technical issue has been solved by using a combined serial-parallel connection of $N$=25 ($N=n^2$, n=5) transmitting-receiving IDTs with 5 groups connected in series and with 5 delay lines in each group. Indeed, if we connect all 25 TR-IDTs parallel, then all the IDTs would reveal a nonzero capacity related to reactive (imaginary) part of admittance at central frequency $f_0$. As a result, the total impedance of 25 IDTs would be reduced significantly to be 3.6 Ω at frequency $f_0$ and would become very different from 50 Ω, the input impedance of a network analyzer. In result, the losses would be gradually increased to be 40 dB.

In contrast, when a combined serial-parallel connection of $N=n^2$ TR-IDTs is used, the capacitance related imaginary part to the total impedance is close to the impedance of a single IDT. More specifically, at the central frequency of $f_0$ for all IDTs the total input impedance is 10 Ω which is close to 50 Ω, the input impedance of network analyzer. As a result of such a connection, the RF signal losses are minor.

For each delay line, the active and reactive parts of the input admittance are given as[20]:

$$ReYin = ReY_{tr} - A, \tag{4}$$

$$ImYin = ImY_{tr} - B \tag{5}$$

where A and B are the terms related to SAW reflection from IDT

When a SAW is reflected from R-IDT, there are two terms which input into the resulting reflection coefficient. Namely, the first term $H$ describes the SAW reflection from electrodes of IDT:[18]

$$ReH_r = g\left[\sum_{k=1}^{M_1}\left[(1-g)^k \cos\left(\frac{2\pi pkf}{f_0}\right)\right]\right] \tag{6}$$

$$ImH_r = g\left[\sum_{k=1}^{M_1}\left[(1-g)^k \sin\left(\frac{2\pi pkf}{f_0}\right)\right]\right] \quad (7)$$

Second term $K_r$ describes the SAW reflection from IDT as a whole[18]:

$$ReK_r = \frac{-Ga_r * Ga_r}{(Ga_r)^2 + [2\pi f(C_r) + Ba_r]^2} \quad (8)$$

$$ImK_r = \frac{Ga_r * (2\pi f C_{Tr} + Ba_r)}{(Ga_r)^2 + [2\pi f(C_{Tr}) + Ba_r]^2} \quad (9)$$

Finally, the SAW reflection coefficient $K_0$ is given by[18,20]:

$$ReK_{0r} = \frac{(ReK_r + ReH_r)}{(1+g)}$$

$$ImK_{0r} = \frac{(ImK_r - ImH_r)}{(1+g)} \quad (9)$$

where $g$ is the reflection coefficient of a single electrode of the R-IDT.
For the *i-th* acoustic channel the terms $A_i$ and $B_i$ are given by:

$$A_i = e^{-\alpha_i L} Y_{12i}\left[(ReK_{0ri})\cos(X_i) + [(ImK_{0ri})\sin(X_i)]\right] \quad (10)$$

$$B_i = e^{-\alpha_i L} Y_{12i}\left[-(ReK_{0ri})\sin(X_i) + [(ImK_{0ri})\cos(X_i)]\right] \quad (11)$$

$$X_i = 2\pi \frac{2L_0 f}{f_{01}}$$

where $a_i L$ is a SAW attenuation in ith acoustic channel. The value of attenuation $a_i L$ is defined by conductivity of $VO_2$ layer an is altered between semiconducting and metallic state by applied electric bias to *i-th* channel (see Figure 1), $L_0$ is the distance between TR-IDT and R-IDT measured in the wavelength at central frequency of the first acoustic channel, L is the length of $VO_2$ layer in SAW channels.
For the device with $N$ SAW delay lines N=25 the total input admittance of one group with 5 delay lines connected parallel is given by:

$$ReYin = \sum_{i=a}^{b} ReY_{TRi} - A_i\,,\; ImYin = \sum_{i=a}^{b} ImY_{TRi} - B_i \quad (12)$$

where ($a$=1, $b$=5), ($a$=6 $b$=10), ($a$=11, $b$=15), ($a$=16, $b$=20), ($a$=21, $b$=25), correspond to formulas for the first, second, third, fourth and fifth group of acoustic channels respectively.
The total impedance $Z_m$ ($m$=1,2…5) of each group of 5 IDTs connected parallel is given by :

$$ReZin_m = \frac{ReYin_m}{(ReYin_m)^2 + (ImYin_m)^2}\,,\; ImZin_m = \frac{-ImYin_m}{(ReYin_m)^2 + (ImYin_m)^2} \quad (13)$$

As these 5 groups are connected in series the impedances are summed and the total input impedance is:

$$Zin = \sum_{m=1}^{5} ReZin_m + j\sum_{m=1}^{5} ImZin_m \quad (14)$$

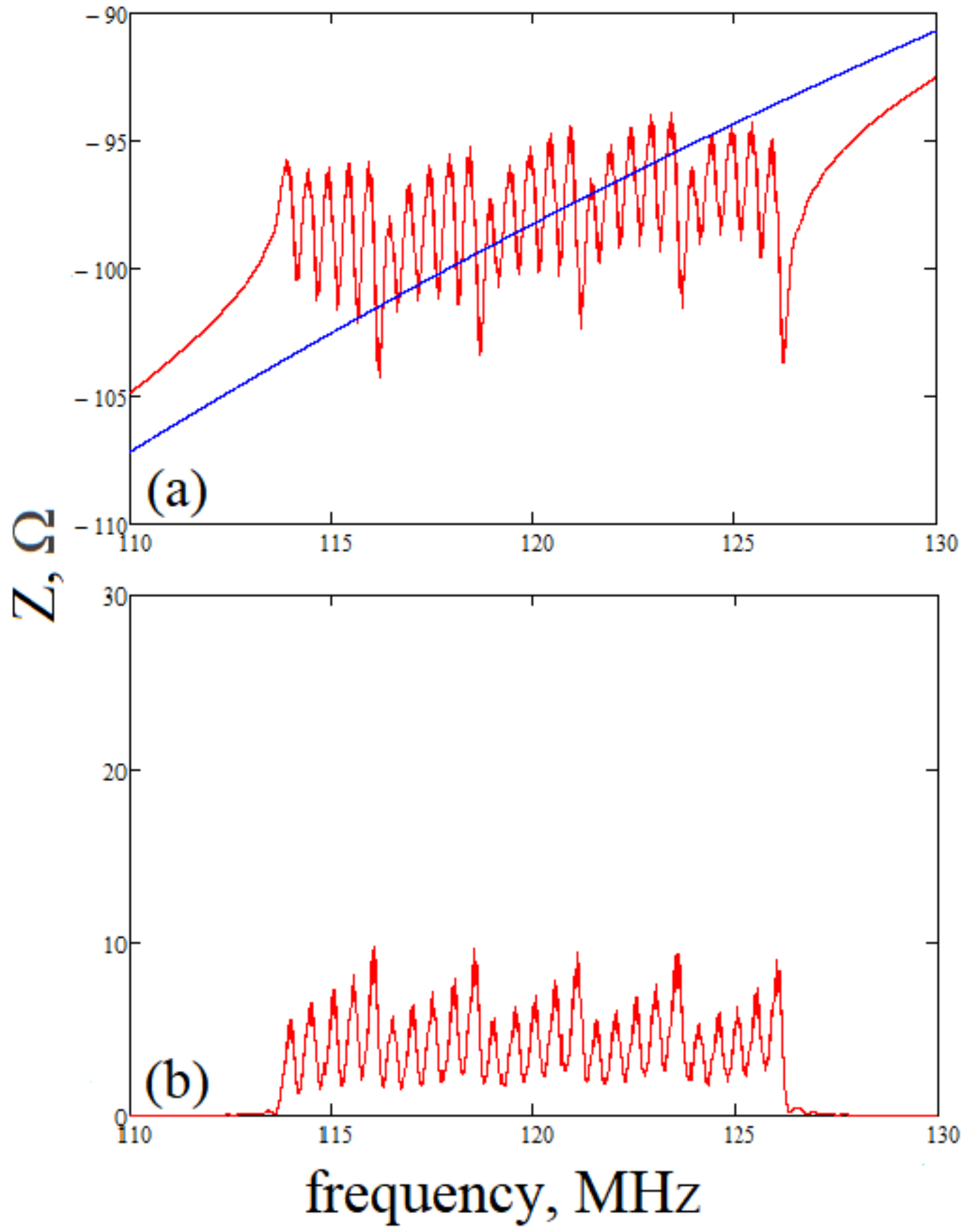

**Fig. 3** Frequency dependence of the device input impedance *Zin*. Imaginary part (a) and real (b) part of impedance. Blue curve in (a) shows the frequency dependence of $1/\omega C_T$

The input impedance of the device with combined serial-parallel connection of IDTs is shown in Figure 5. Outside the transmission band the imaginary part of the device impedance limits to capacity related input to impedance $1/\omega C_T$ as shown in Figure 5 (a). This capacity is equal to capacity of TR-IDT.

*Work of the multi-channel SAW device*

As shown in Figure 1 the device input impedance is a shoulder of the differential bridge circuit based on symmetrical transformer. Another shoulder is a capacity of transmitting-receiving IDT. The differential bridge schematic is described by the matrix *A*:

$$A = \begin{bmatrix} \frac{Z_1+Z_2}{Z_1-Z_2} & \frac{2 \cdot Z_1 \cdot Z_2}{Z_1-Z_2} \\ \frac{2}{Z_1-Z_2} & \frac{Z_1+Z_2}{Z_1-Z_2} \end{bmatrix} = \begin{bmatrix} a_{11} & a_{12} \\ a_{21} & a_{22} \end{bmatrix} \tag{16}$$

where $Z_1=2Z_{in}$, $Z2=Z$ (see Figure 1 and ref. 21)
The transmission coefficient of the bridge is defined as[21] :

$$S_{21} = \frac{1}{a_{11}+\frac{a_{12}}{R_L}+a_{21}R_g+\frac{a_{22}}{R_L}R_g} \tag{17}$$

where $R_G$ and $R_L$ the resistance of generator and load respectively.
Assume that $R_G$=$R_L$=$R$. Then, after simple transformations, the transmission coefficient S21 is defined as:

$$S_{21} = \frac{Z_1 - Z_2}{Z_1 + Z_2 + R + \frac{Z_1 Z_2}{R}} \qquad (18)$$

The frequency dependence of the transmission coefficient $K$ is shown in Figure 4 (a). As one can notice the RF losses are about 38 dB in the whole frequency range. Indeed, the reactive (imaginary) part of the input impedance is much greater than active (real) part which results in significant losses.

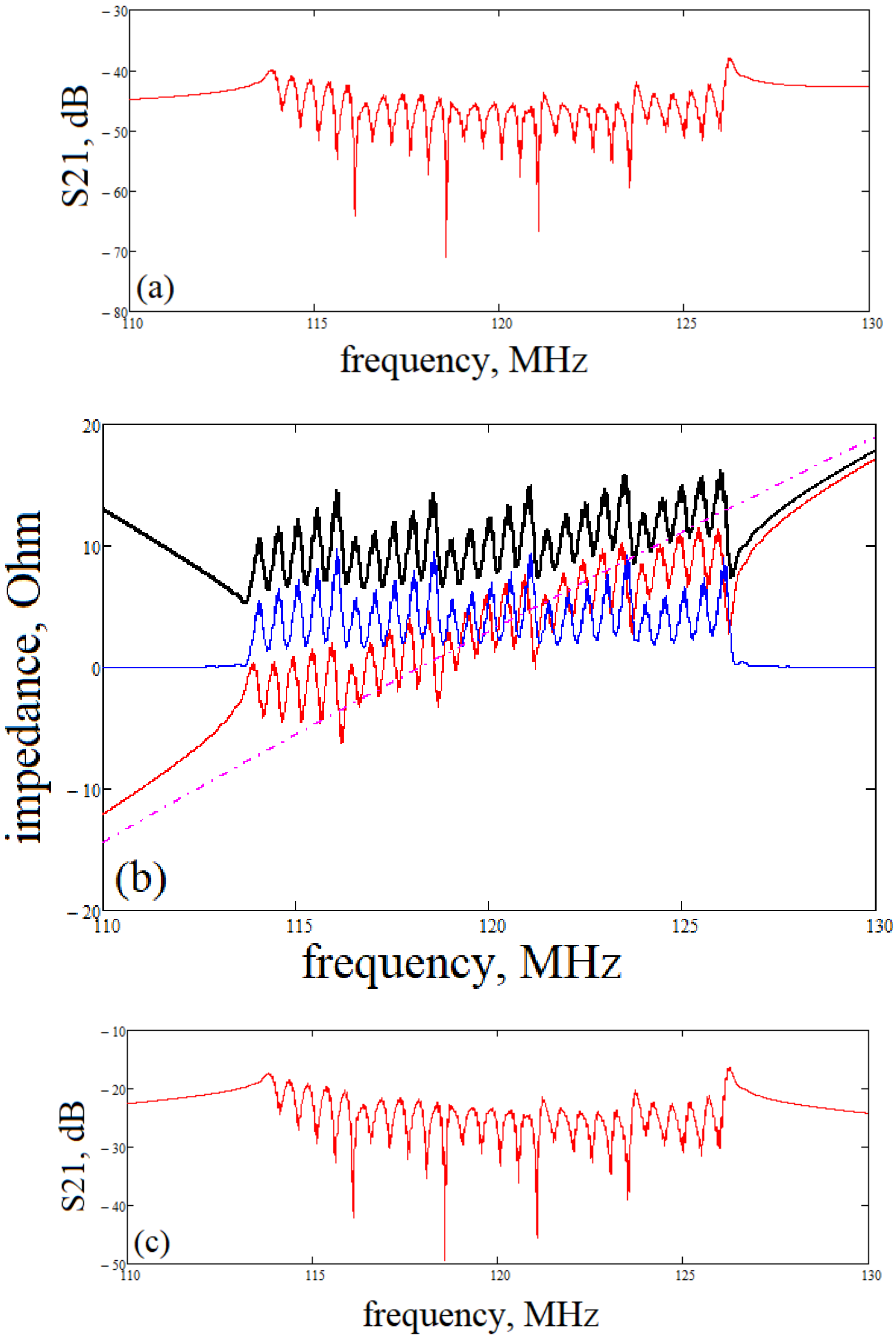

**Fig. 4** S21 characteristics of the differential bridge in the electric circuit of the SAW device (see Figure 1) before (a) and after (c) the bridge circuit matching using additional inductance.The impedance of the device (black), imaginary part (red) and a modulus of the active (real) part (blue) of the impedance. The capacitance related imaginary part of the (dotted line) (b).

We used a sequential inductance *L1* to reduce losses (See Figure 1). The inductance was chosen to reduce by zero the imaginary part of the impedance at the middle of the working frequency range as shown with the red curve in Figure 6 (b). At other frequencies the total impedance of the device (black curve) is nearly coincident with the modulus of the active (real) part (blue curve). The matching of the bridge scheme by using an additional inductance results in reduced losses and increased RF signal as shown in Figure 6 (c).

The proposed SAW device allows us to in-situ monitor the SAW attenuation as well as the phase shifts[17] in all 25 acoustic channels. By these means the resistance as well as the temperature[17] alterations in all the lines of an active $VO_2$ based THz grating are monitored. Indeed, as we showed recently[17] the altered electric carrier density in $VO_2$ layer due to MIT results in change of the attenuation of a respective SAW from 40 dB to 20 dB. Also, as the SAW velocity as well as phase of a SAW passing through the $LiNbO_3$ substrate is linearly dependent on local surface temperature of acoustic channel[17], that unique property can be also effectively exploited for monitoring of a local temperature distribution on the surface of active $VO_2$ based THz grating which lines are controlled by local heating. Important, that monitoring of local temperature and carrier density in $VO_2$ film located onto acoustic channel are nearly independent processes even within hysteresis loop of MIT.

## 3 Results and discussion

Several options to configure a multi-channel SAW device to in-situ monitor an active THz grating are discussed below.

*Binary reflecting amplitude grating with adjusted period*

The resistance of $VO_2$ layer is programmed to have levels called “0” and “1” which correspond to the semiconductor and metallic states of VO2 before and after MIT, correspondingly. A normalized SAW attenuation in channels *n* are called in similar manner as $\alpha_n \times L = \{1, 1, 0, 0, 1, 1, 0, 0, 1, 1 \ldots\}$. This case corresponds to a binary reflecting amplitude grating in which the reflection of THz waves from $VO_2$-based lines has two values called “0” and “1”. The period of the grating shown in Figure 4(a) is equal to width of “pixels”, i.e. to the two widths of acoustic channels.

When the resistance of $VO_2$ layer in acoustic channels *n* is configured to set the corresponding normalized SAW attenuations as $\alpha_n \times L = \{1, 0, 1, 0, 1, 0, 1, 0, 1, 0 \ldots\}$, then the period of the grating is equal to the width of a single acoustic channel. The corresponding response of the SAW device is shown in Figure 4(b).

*Gradient type reflecting amplitude grating*

Resistance of $VO_2$ layer in the first 10 acoustic channels is adjusted to have increasing values which correspond to normalized SAW attenuations of $\alpha_n \times L = 0.1 \times n$, where $n=1\text{-}10$. The S21 response of a SAW device in such configuration is shown in Figure 4(c).

This case corresponds to reflection type amplitude grating in which the front of THz wave reflected from the $VO_2$-based lines (pixels) is adjusted by electric currents to have several different levels. Sine grating is a most known example of more general class of gradient type gratings.

This advanced option becomes possible when using unique properties of a composite $VO_2/TiO_2/LiNbO_3$ recently reported by us.[17] As we recently reported, the resistance of $VO_2$ in such composites is adjusted to any state between semiconducting and metallic ones allowing tuning THz reflection as well.

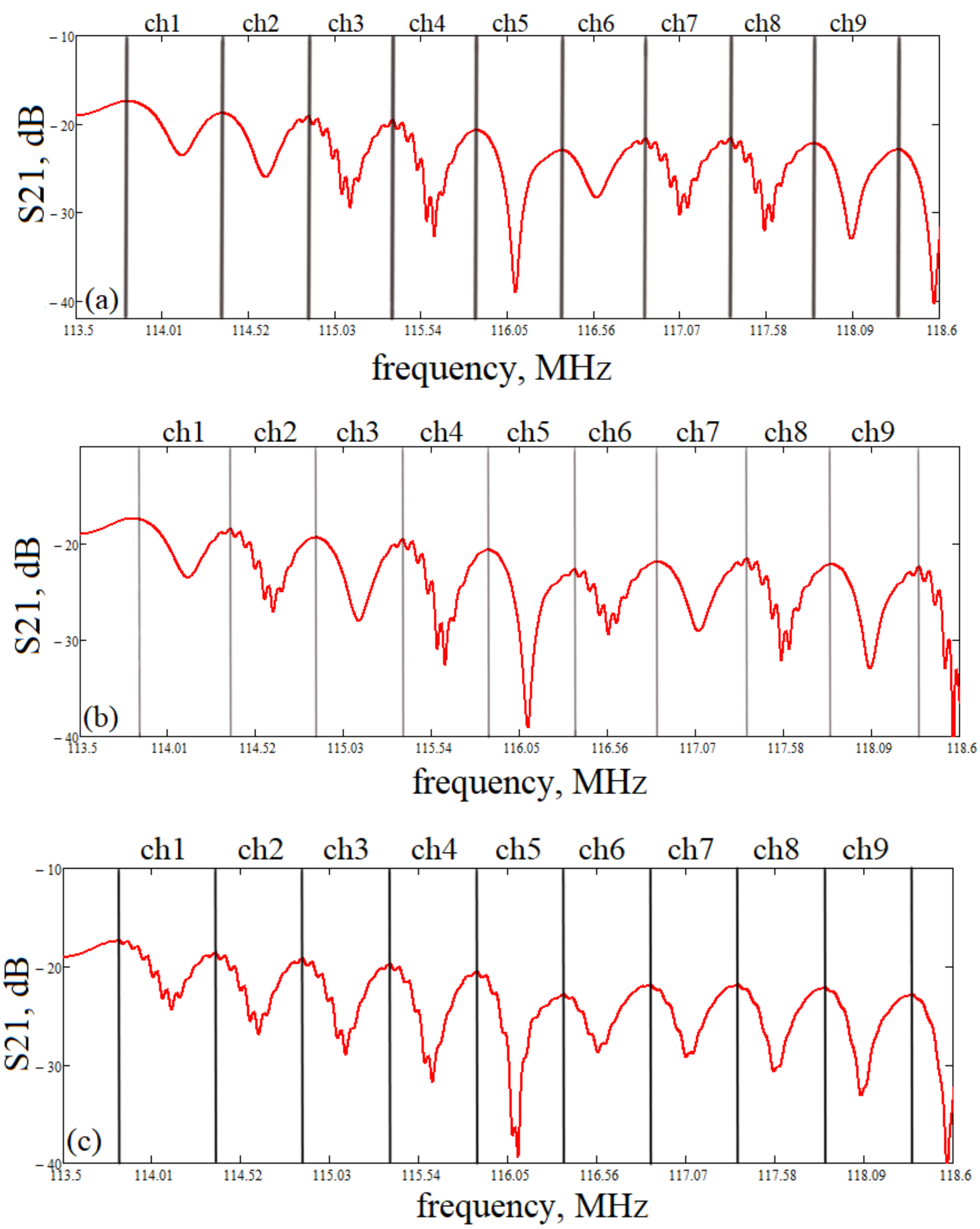

**Fig. 5** S21 characteristics of multichannel SAW device.RF responses correspond to different options of device programming. Options of configuring as binary amplitude grating with the altered period: channels are coded as [0011001100] (a) and as [0101010101] (b). “0” and “1” correspond to the semiconductor and metallic states of $VO_2$ film before and after MIT. Option of configuring as gradient type amplitude grating with SAW attenuations in channels of $\alpha_n L$=0.1×n, where n=1-10 (c).

The proposed method of in-situ monitoring of the active THz grating is beneficiary over a simple approach of applying currents to $VO_2$ based elements of a metasurface without any feedback. Indeed, some meta-atoms or even larger groups (pixels) may be damaged with time. Also, some undesired defects may be appeared during lithographic fabrication of meta-atoms or electrodes. Simple controlling of the resistance of the lines of the grating with many meta-atoms that are electrically connected parallel is not always a reliable strategy. In contrast, controlling of the attenuation of a SAW passing through the meta-atoms made of $VO_2$ allows one in-situ monitor all the minor alterations of electric carriers in VO2 as well as a local temperature on the surface of the device.[17] A reliable and guaranteed controlling of the THz reflection coefficients within each single pixel is achieved this way. Also, in some applications they require the option of a guaranteed programming of distribution of the local THz reflections within metasurface even before the THz radiation is incident to the device. The proposed multi-channel SAW monitoring method allows such guaranteed configuring.

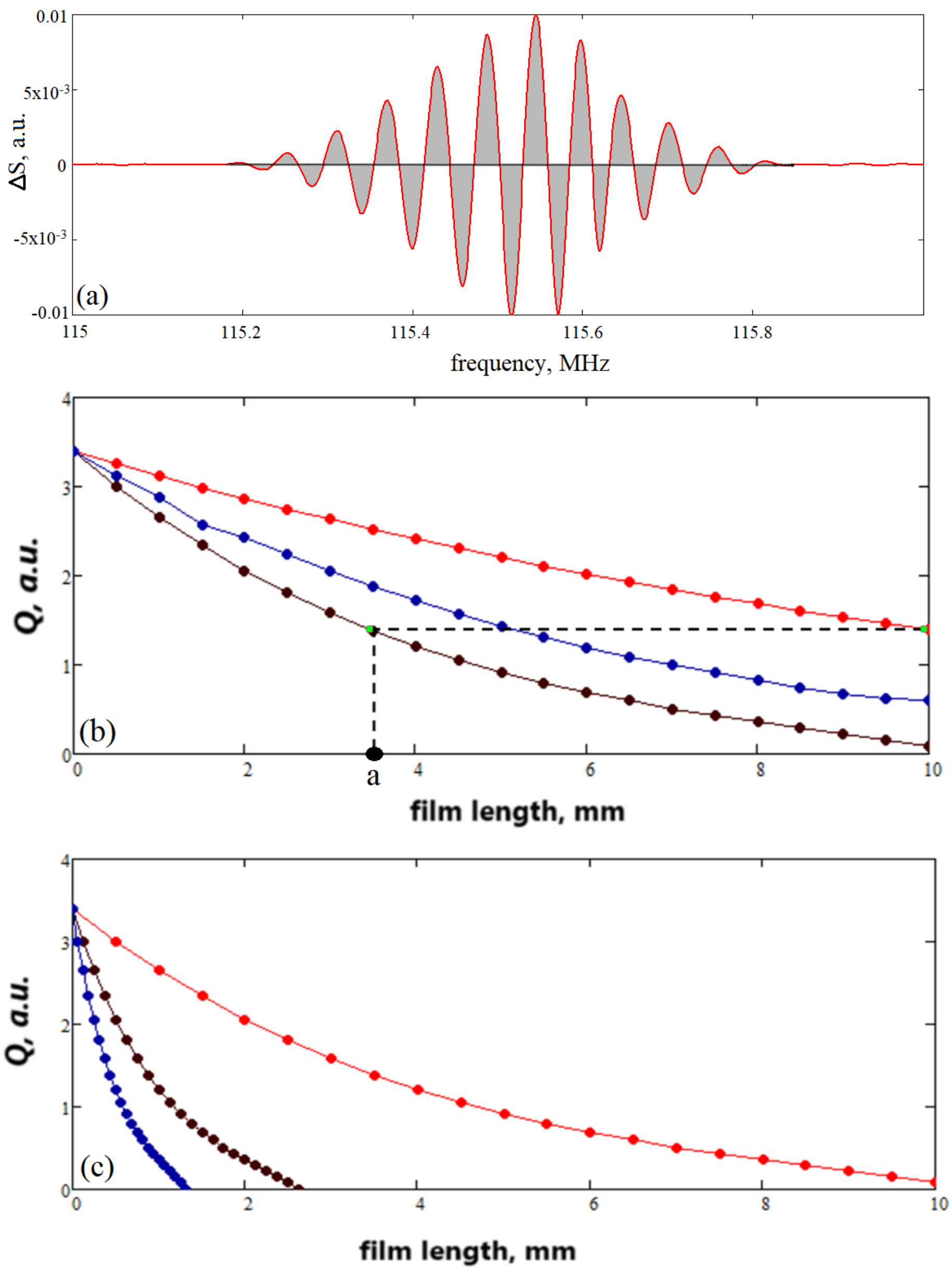

**Fig. 6** $\Delta S$=*|S21(oscillating)-S21(smooth)|* for SAW in a single acoustic channel. *S21(oscillating)* – oscillating curve, when acoustic losses are present. S21(smooth) - smooth S21 response corresponding to a minor attenuation in acoustic channel (metallic state of $VO_2$) (a). Results of numerical modeling for SAW at 120MHz: The area $Q$ under curves difference ($\Delta S$) calculated for different lengths of the SAW propagation via $VO_2$ film. Note that shown device responses correspond to the length of acoustic channel with VO2 film and the SAW is propagating twice in accordance with device construction. The case with $VO_2$ film in a metallic state when SAW losses are minor (red), $VO_2$ is partially switched (blue) and $VO_2$ is in semiconducting state when the losses are significant (brown) (b). $Q$ as function of SAW propagation lengths via the $VO_2$ film at SAW frequencies of 120 MHz (red), 433 MHz (brown) and 920 MHz (blue) (c).

The acoustic attenuation does limit the propagation length of a SAW due to electro-acoustic interaction of the wave with carriers in $VO_2$ layer. When the $VO_2$ layer is in metallic state the SAW losses are minor and amplitude of oscillations is significantly increased when the $VO_2$ layer becomes semiconducting. We found it handy to use the value of $\Delta S$=|S21(oscillating)-S21(smooth)| (see Figure 6 (a)) , the difference between areas under the oscillating S21 response when acoustic losses a nonzero and smooth S21 response when SAW attenuation is minor. Indeed, $\Delta S$ can be used as feedback information on the current level of the losses in the acoustic channels.

In particular, by analyzing the $\Delta S$ values for the different lengths of the acoustic channel with $VO_2$ film one may define the maximal possible length of the acoustic channel when the feedback signal is still detectable. The modeled propagation distances of SAW at 120 MHz via $VO_2$ film are shown in Figure 6 (b). Indeed, the SAW is able to propagate twice via 10 mm acoustic channel with $VO_2$ film when it is in metallic state (see red curve in Figure 6 (b)). However, when $VO_2$ film is in semiconducting state, the acoustic losses are increased and the maximal possible length of the acoustic channel with $VO_2$ film is reduced to 3.5 mm (see brown curve, point *a* and dotted lines in Figure 6 (b)). Indeed, after passing twice through the acoustic channel longer than 3.5 mm the losses due to electro-acoustic interaction of a SAW with carriers in the $VO_2$ layer become so significant that a proposed monitoring method can not be applied any more at this SAW frequency.

Next, we model the maximal possible acoustic channel length when the device is designed at frequencies of 120 MHz, 433 MHz and 920 MHz. As shown in Figure 6 (c), at frequency of 433 MHz a feedback signal is still detectable when the acoustic channel length is shorter than ~2 mm and the SAW is completely attenuated in longer channel with $VO_2$ film in semiconducting state. At frequency of 920 MHz the SAW can be detected only in acoustic channel with VO2 film of the length shorter than ~1 mm (see the blue curve in Figure 6 (c)).

The proposed method is general and can be used to monitor many different types of metasurfaces. In discussed above device at 120 MHz with continuous $VO_2$ film covering whole the acoustic channel the SAW attenuation is 15 and 45 dB for semiconducting and metallic states respectively. In contrast to amplitude type metasurface with the continuous active layer in phase type metasurface with $VO_2$ based split-ring resonators may cover only a small area of the acoustic channel. The SAW attenuation depends on the area of the acoustic channel but not only of its length. When $VO_2$ is in semiconducting state attenuation of acoustic waves will be smaller compared to the case of the continuous film of the same length. The SAW S21 response will reveal larger oscillations as a result which will enhance the signal-to-noise ratio and measurement dynamic range of the method.

## 4 Conclusions

In summary, we proposed the method for in-situ monitoring of the states of all the lines of the $VO_2$ based active THz amplitude metagrating. The feedback information on the SAW attenuation in acoustic channels, which coincide with lines of the $VO_2$ grating, is coded in a single RF response with *N* independent frequency channels. This way the MIT related resistance alteration in all pixels of the active $VO_2$ based grating is monitored in advance even without incident THz radiation. The approach allows one obtain the guaranteed controlling of pixels switching as well as detect aging or false switching related to technological defects. The scenarios of programming and monitoring of a binary reflecting amplitude type grating with adjusted period and gradient type reflecting amplitude grating are theoretically examined. Also, the model has revealed the constraints which have to be applied to the length of the acoustic channel when designing the active part of the THz grating. For the metagrating with continuous $VO_2$ film in acoustic channel the feedback signal is detectable in acoustic channel shorter than

~6-10 mm in device designed for 120 MHz acoustic waves and it has to be reduced to 1 mm at 920 MHz.

**Author contributions** G.Y.K. prepared the original manuscript and developed the design of SAW device. G.Y.K. and M.V.O. performed the theoretical modeling and figures preparation. V.E.K. suggested the concept of the method to monitor THz metasurface using SAW. V.E.K. made formal analysis and revised the manuscript.

**Funding** The work was supported by Russian Science Foundation, grant № 24-29-00175 at SFU

**Data availability** Data will be made available on request.

## Declarations

**Conflict of interest** The authors declare no competing interests.